\documentclass[twocolumn,tight,dvipsnames]{aastex63}
\usepackage{amsmath,mathtools,mathrsfs,bm}
\usepackage{hyperref}
\usepackage{verbatim}
\usepackage{xspace}

\usepackage{xcolor}

\def\muas{\mu{\rm as}} 
\def\uas{$\mu$as\xspace} 
\def\dd{\mathrm{d}}
\def\m87{M~87}
\def\Dc{\mathcal{D}}

\newcommand{\VIDA}{\texttt{VIDA}\xspace}
\newcommand{\VIDAjl}{\texttt{VIDA.jl}\xspace}
\newcommand{\ReX}{\texttt{REx}\xspace}
\newcommand{\ehtim}{\texttt{eht-imaging}\xspace}

\newcommand{\C}{\texttt{C}\xspace}

\newcommand{\norm}[1]{\left\lVert#1\right\rVert}

\renewcommand{\L}{{\mathcal L}}

\defcitealias{EHTCI}{Paper~I}
\defcitealias{EHTCII}{Paper~II}
\defcitealias{EHTCIII}{Paper~III}
\defcitealias{EHTCIV}{Paper~IV}
\defcitealias{EHTCV}{Paper~V}
\defcitealias{EHTCVI}{Paper~VI}

\shorttitle{\VIDA}
\shortauthors{Tiede et al.}

\begin{document}
  
\title{Variational Image Feature Extraction for the Event Horizon Telescope}


\correspondingauthor{Paul Tiede}
\email{paul.tiede@cfa.harvard.edu}

\author[0000-0003-3826-5648]{Paul Tiede}
\affiliation{ Perimeter Institute for Theoretical Physics, 31 Caroline Street North, Waterloo, ON, N2L 2Y5, Canada}
\affiliation{ Department of Physics and Astronomy, University of Waterloo, 200 University Avenue West, Waterloo, ON, N2L 3G1, Canada}
\affiliation{ Waterloo Centre for Astrophysics, University of Waterloo, Waterloo, ON N2L 3G1 Canada}
\affiliation{ Black Hole Initiative at Harvard University, 20 Garden Street, Cambridge, MA 02138, USA}
\affiliation{ Center for Astrophysics | Harvard \& Smithsonian, 60 Garden Street, Cambridge, MA 02138, USA}

\author[0000-0002-3351-760X]{Avery E. Broderick}
\affiliation{ Perimeter Institute for Theoretical Physics, 31 Caroline Street North, Waterloo, ON, N2L 2Y5, Canada}
\affiliation{ Department of Physics and Astronomy, University of Waterloo, 200 University Avenue West, Waterloo, ON, N2L 3G1, Canada}
\affiliation{ Waterloo Centre for Astrophysics, University of Waterloo, Waterloo, ON N2L 3G1 Canada}

\author[0000-0002-7179-3816]{Daniel C. M. Palumbo}
\affiliation{ Black Hole Initiative at Harvard University, 20 Garden Street, Cambridge, MA 02138, USA}
\affiliation{ Center for Astrophysics | Harvard \& Smithsonian, 60 Garden Street, Cambridge, MA 02138, USA}

\begin{abstract}

Imaging algorithms form powerful analysis tools for VLBI data analysis. However, these tools cannot measure certain image features (e.g., ring diameter) by their non-parametric nature. This is unfortunate since these image features are often related to astrophysically relevant quantities such as black hole mass. This paper details a new general image feature extraction technique that applies to a wide variety of VLBI image reconstructions called \textit{variational image domain analysis}. Unlike previous tools, variational image domain analysis can be applied to any image reconstruction regardless of its structure. To demonstrate its flexibility, we analyze thousands of reconstructions from previous EHT synthetic datasets and recover image features such as diameter, orientation, and ellipticity. By measuring these features, our technique can help extract astrophysically relevant quantities such as the mass and orientation of the central black hole in M 87.
\end{abstract}


\section{Introduction}\label{sec:intro}
\nocite{EHTCI,EHTCII,EHTCIII,EHTCIV,EHTCVI}

Generating quantitative measurements about intrinsic radio images from very long baseline interferometry (VLBI) is a computationally and theoretically difficult task. Even in principle, it is impossible to associate a single unique image with a given set of visibility data.  In practice, the small number of stations participating in the Event Horizon Telescope (EHT), and thus sparse coverage in the $u$-$v$ plane, results in variety of potential image structures.  As a result, the process of reaching quantitative conclusions about image features requires significant additional analysis. For the EHT analysis of the 2017 \m87 data, this has taken two, complementary forms\citep[][]{EHTCI}.

The first is a traditional Bayesian parametric modeling approach \citep[][hereafter \citetalias{EHTCVI}]{EHTCVI}.  Simple geometric models are fit to the visibility data, and direct quantitative inferences about the image properties encoded within the model, e.g., diameter, are possible.  Higher model fidelity improves the accuracy of the derived measurements, and thus this requires an input assumption that the key model features capture the ``truth''.

The second is a nonparametric approach, usually referred to as ``imaging'' \citep[][hereafter \citetalias{EHTCIV}]{EHTCIV}.  This category of methods is broadly defined, and includes deconvolution algorithms like CLEAN \citep{hogbom_aperture_1974, schwarz_mathematical-statistical_1978, Clark_1980, Schwab_1984} and forward modeling approaches like ``maximum entropy'' \citep[][]{frieden_restoring_1972, gull_image_1978, narayan_maximum_1986}, regularized maximum likelihood (RML) analyses \citep{Chael_2016, Chael_2018, Akiyama_2017a, Akiyama_2017b}, and Bayesian imaging \citep{Broderick_2020b}. The output of imaging by the EHT has been an ensemble of image reconstructions that reproduce the observed visibility data \citealt{EHTCIV}. The imaging methods have the significant advantage that they are extremely flexible, and therefore are reasonably expected to cover the ``truth''.  However, unlike parametric modeling, imaging methods do not give direct quantitative measurements of the image features of interest, e.g., ring diameter, width, orientation, etc.  Therefore, reaching quantitative conclusions about the image properties requires an additional processing step, which we call ``feature extraction''. It is this final step that is the subject of this paper.

Feature extraction is similar to the Bayesian parametric modeling applied in the image, rather than visibility, domain.  However, it differs in one important respect: by virtue of being performed {\em after} imaging, the class of applicable ``models'', i.e., image features to be measured, is already well known.
For example, in \citetalias{EHTCIV} and \citetalias{EHTCVI}, quantitative image features were extracted by the algorithm ``ring extractor'' \ReX \citep{chael_simulating_2019}. However, \ReX is only applicable to images that have a dominant ring-like feature. In general, images from the EHT can have a complex structure and are dependent on the intrinsic source. The active galactic nuclei 3C~279 is displays a jet morphology which while poorly described by a ring, can be described by a set of Gaussians \citep{3c279EHT}.

One possible approach to feature extraction is to ``template'' relevant image features using a transformation. For example, the Hough transform \citep{Hough1964, duda}, is used to extract rings and other shapes from images using template matching. A related method is to approximate the complicated image reconstruction with parametric templates that describe the features of interest. This idea is more akin to the parametric/geometric modeling approach of visibility data in \citetalias{EHTCVI} and is the approach taken in this paper.

Any comparison requires a suitable quality metric, i.e., objective function.  Because the total flux is typically an arbitrary rescaling and the image brightness is positive definite, there is a natural identification between the flux-normalized image and probability distribution.  This motivates the use of ``divergences'' as an extremely flexible class of objective functions for comparing images; for this reason divergences have been used extensively in image processing  \citep[e.g.,][]{goudail_bhattacharyya_2004, choi_feature_2003, Aherne1998}.

We adopt the method similar to variational inference \citep{blei_variational_2017}, in which complicated distributions are approximated by simple parametric forms, with parameters estimated via the minimization of an appropriate divergence.  In this paper, we develop a number of appropriate parametric forms and explore the performance of a variety of divergence for application to image feature extraction. Therefore, we call this method \textit{variational image domain analysis}, or \VIDA.  The \VIDA algorithm has been implemented in the open source package \VIDAjl\footnote{\url{https://github.com/ptiede/VIDA.jl}} written in Julia \citep{julia}.

There are two main reasons that imaging reconstruction followed by feature extraction is advantageous to directly fitting simple geometric models to the data. First, choosing the correct geometric model is difficult before imaging, meaning imaging is the first step in both methods. Second, fitting simple geometric models can ``underfit'' the data leading to biased results. For instance, in \citetalias{EHTCVI} additional ``nuisance'' Gaussians were required to obtain a reasonable reduced chi-square. On the other hand, Bayesian imaging techniques (e.g., \citep{Broderick_2020b}) make minimal assumptions about source structure. Combining the Bayesian imaging posterior with \VIDA then provides a deterministic map from image to feature posteriors. \VIDA's feature posteriors can then be compared to the geometric modeling results testing whether the features are robust across methods. Therefore, \VIDA fills a gap in the EHT modeling pipeline and is generic, unlike current EHT tools.

The layout of the paper is as follows:
In \autoref{sec:VIDA}, we present the details of \VIDA. Different types of templates implemented, and the objective function used to find the best approximation to the true image are detailed. \autoref{sec:validation} applies \VIDA to a variety of ring-like image reconstructions from the test set of \citetalias{EHTCIV} and compares the results to \ReX. This is an empirical demonstration that we can recover the optimal template, even though the objective function is non-convex. In \autoref{section:asym}, we demonstrate \VIDA's flexibility by applying it to non-ring images from the test set of \citetalias{EHTCIV}. Finally, the conclusions are detailed in \autoref{sec:conclusions}. Additional material related to the model and results for CLEAN is presented in Appendices \ref{appendix:rex} -- \ref{appendix:clean}.

\section{Variational Image Domain Analysis}\label{sec:VIDA}
The critical insight behind \VIDA is that images (sans polarization and modulo total flux) and probability densities are in one-to-one correspondence. Namely, images are point-wise positive and integrable. Probability divergences are a natural class of objective functions used to compare two probability distributions. Furthermore, they have been used in image feature extraction and similarity measures before. Divergences thus form a natural objective function between the image reconstruction and the template. 
\VIDA consists of three ingredients:
\begin{enumerate}
    \item \textbf{Image} $I(\alpha,\beta)$ whose features we want to extract
    \item \textbf{Template} (or approximate image) that parameterizes the features of interest, e.g. ring radius 
    \item A \textbf{divergence}, i.e. the objective function we minimize
\end{enumerate}
Each of these building blocks have independent abstract types in \VIDAjl enabling users to easily implement additional images, templates, and divergences\footnote{See the documentation at \url{https://ptiede.github.io/VIDA.jl/dev/} for a tutorial on how to add additional templates}.
In \autoref{ssection:template} and \ref{ssection:divergences} we will review the templates and divergences currently implemented in \VIDA respectively.

\subsection{Image Templates in \VIDAjl}\label{ssection:template}
The choice of template used will depend on the structure of the image. For example, the images of \m87 from \citetalias{EHTCIV} are ring-like, while the reconstructions of 3C~279 from \citet{3c279EHT} can be described by several Gaussian brightness distributions. In this section, we present the various templates that are implemented in \VIDAjl.

\subsubsection{Gaussian template}
To model a source of compact flux we include an asymmetric Gaussian template. The parameters of the Gaussian template are:
\begin{enumerate}
    \item The size, $\sigma = \sqrt{\sigma_a\sigma_b}$, where $\sigma^2_{a,b}$ are the variances in the principal directions of the Gaussian.
    \item The ellipticity, $\tau = 1-\sigma_b/\sigma_a$, measures the ellipticity of the Gaussian and we assume $\sigma_a>\sigma_b$.
    \item $\xi$, rotation angle (relative to the Gaussian center) of the principal axes measured east of north.
    \item $x_0,\, y_0$, the center of the Gaussian.
\end{enumerate}

\begin{figure*}[!t]
    \centering
    \includegraphics[width=\linewidth]{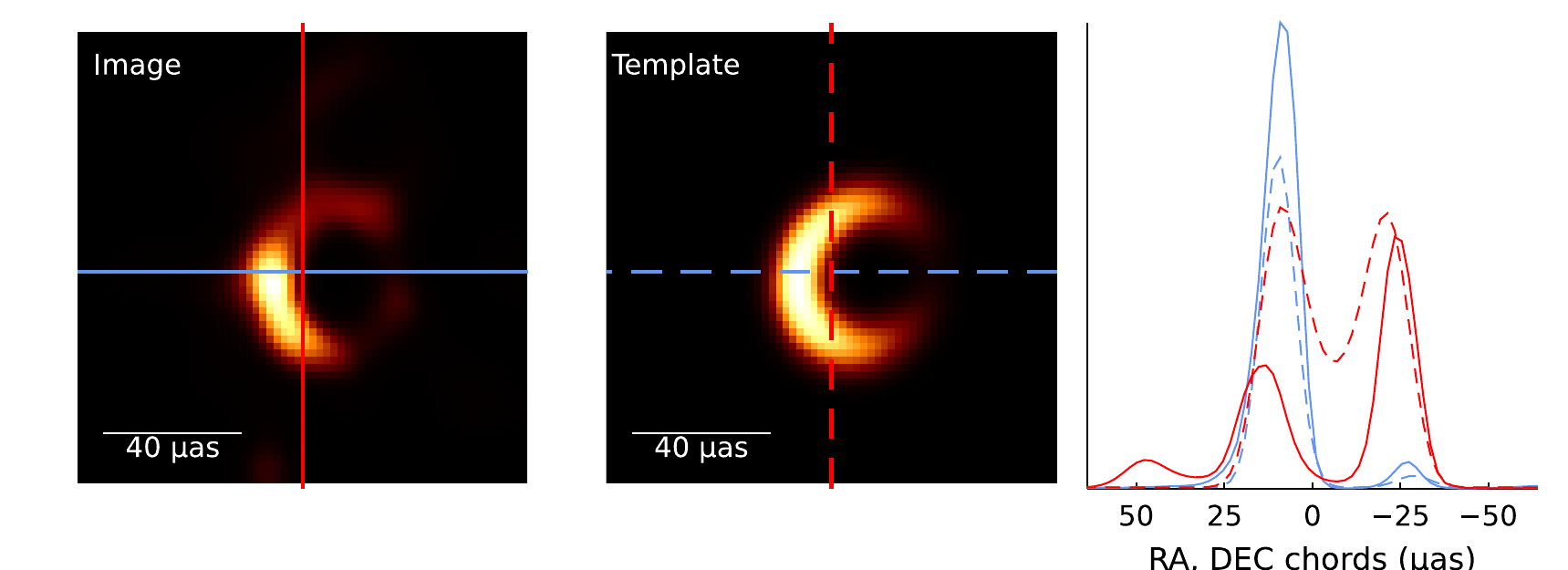}
    \caption{An example of \VIDA run. \textit{Left:} image reconstruction of a GRMHD simulation from \citetalias{EHTCV} and \citetalias{EHTCIV}. \textit{Middle:} \VIDA reconstruction using the \texttt{GeneralGaussiantemplate} and Bh divergence. \textit{Right:} vertical (red) and horizontal (blue) chords through through the center of light of the truth image. The dashed lines are for the \VIDA optimal template, and solid for the reconstruction. This plot can be made in the \VIDAjl package using the \texttt{triptic} function and is used to assess the quality of the template approximation.}
    \label{fig:eccentricity}
\end{figure*}

\subsubsection{Disk template}
In \autoref{sec:validation} we will test \VIDAjl on a number of synthetic data tests. One of the test images is a disk. To approximate disks we use the template:
\begin{equation}\label{eq:disk_template}
    f_{\rm Disk}(r; r_0,\alpha,x_0,y_0) = N
    \begin{cases}
    1 & r<r_0 \\
    \exp(-(r-r_0)^2/2\alpha^2) & r > r_0,
    \end{cases}
\end{equation}
where $r_0$ is the radius of the flat disk, $\alpha$ controls the smoothness of fall off and $N$ ensures the template is normalized. The radial distance, $r$, is relative to the center $(x_0,\, y_0)$. When $r_0=0$ this template reduces to a symmetric Gaussian with standard deviation $\alpha$.

\subsubsection{Ring templates}
One of the principal quantities of interest in images of \m87 is the ring diameter, $d_0$, since it is related to the mass of the central black hole. Additionally, the ring is expected to have some thickness, $w$, due to the emitting material around the black hole. The simplest template would be a circular Gaussian ring with some thickness. Doppler boosting however can cause the emission to appear asymmetric. To model this a slash can be added to the ring template to include the brightness asymmetry. Additionally, the ring itself does not have to be circular. Ring ellipticity could occur from e.g., the emitting material not being azimuthally symmetric around the black hole. As well, due to the sparse coverage of the EHT array and imaging algorithms, ellipticity may be introduced into the reconstructions. Consolidating each of these features into a template parameter, we get the following:
\begin{itemize}
    \item $d_0$: the geometric mean of the semi-major, $a$, and semi-minor, $b$, axis $d_0 = 2\sqrt{ab}$ which is related to the area of the ellipse, $\pi (d_0/2)^2$.
    \item $\tau$: the ellipticity of the ellipse, $\tau = 1-b/a$
    \item $\xi_\tau$: the position angle of the semi-major axis measured north of east.
    \item $w$: the width of the Gaussian ring, defined to be the full width half max (FWHM) of the Gaussian, i.e. $w = 2\sqrt{2\log 2}\sigma$, where $\sigma$ is the standard deviation.
    \item $s$: the strength of the slash described in equation \eqref{eq:slash}.
    \item $\xi_s$: the position angle of the slash measured east of north.
    \item $(x_0,y_0)$: the center of the ring.
\end{itemize}
The functional form of the template is given by:
\begin{equation}\label{eq:templateFunction}
    h_{\theta}(x,y) = S(x,y; s,\xi)\exp\left[ - \frac{(d_\theta(x, y))^2}{2 \sigma^2} \right], 
\end{equation}
where $S(x,y;s,\xi)$ is the slash function and $d_\theta(x,y)$ is the minimum distance between the ellipse with parameters $\theta=d_0,\, \tau,\, x_0,\, y_0$ and the point $x,\,y$. If $\tau=0$, $d_{\tau=0}(x, y) = \left|\;\norm{\mathbf{x}-\mathbf{x}_0} - d_0/2\right|$. However, for an ellipse there is no analytical equation and instead once has to numerically minimize the function,
\begin{equation}
        L(x,y,e_x,e_y) = \norm{(x-e_x,y-e_y)},
\end{equation}
subject to the constraint that $e_x,e_y$ are points on the ellipse with parameters $d_0, \tau, \xi_\tau$.

For the slash function $S$, we use a first order cosine expansion in azimuthal angle $\phi$ around the center $x_0,\,y_0$:
\begin{equation}\label{eq:slash}
    S(x,y; s,\xi_s) = N_0\left[1+s\cos(\phi-\xi_s)\right],
\end{equation}
where $N_0$ is a normalization factor to ensure the template is unit normalized. We restrict $s\in[0,1]$ to prevent image brightness from becoming negative. In the \VIDAjl package, this template is called \texttt{GeneralGaussianRing} (GGR), and an example reconstruction using said template is shown in \autoref{fig:eccentricity}. 

Additionally, \VIDAjl has a number of other ring-like templates currently implemented: 
\begin{itemize}
    \item \texttt{GaussianRing}: Symmetric Gaussian ring with constant azimuthal intensity (i.e. GGR with $\tau,s=0$)
    \item \texttt{SlashedGaussianRing}: Symmetric Gaussian ring with azimuthal slash described by \autoref{eq:slash} (i.e. GGR with $\tau = 0$)
    \item \texttt{EllipticalGaussianRing}: Elliptical Gaussian ring with constant azimuthal brightness (i.e. GGR with $s = 0$)
    \item \texttt{TIDAGaussianRing}: GGR template where the slash and ellipticity position angle are either aligned or anti-aligned.
\end{itemize}
We also include a more general version of the GGR called the \texttt{CosineRing\{N,M\}}. This template is similar to the GGR template but where the width, $\sigma$, and slash function \eqref{eq:slash} are replaced by a higher order cosine expansion in azimuthal angle $\phi$:
\begin{align}
    S_{M}(\phi; \mathbf{s}, \bm{\xi}^{(s)}) &= 1 - \sum_{m=1}^Ms_m\cos\left[m(\phi-\xi^{(s)}_{m})\right],\label{eq:cos_slash}\\
    \sigma_{N}(\phi; \bm{\sigma}, \bm{\xi}^{(\sigma)}) &= \sigma_0 + \sum_{n=1}^{N}\sigma_n\cos\left[n(\phi-\xi^{(\sigma)}_n)\right],\label{eq:cos_width}
\end{align}
where $\mathbf{s},\, \bm{\sigma},\, \bm{\xi}^{(.)},$, are vectors with the cosine expansion coefficients of the slash, standard deviation, and angular offset\footnote{In this paper, bold characters will represent vectors}. We can reproduce the \texttt{GeneralGaussianRing} template by setting $M=1$ and $N=0$ in \eqref{eq:cos_slash} and \eqref{eq:cos_width} respectively. This template can be used if the image has a ring-like feature that has a bumpy azimuthal profile.

\subsubsection{Constant Template}
We found that many image reconstructions had a diffuse intensity throughout the image due to poor dynamic range from sparse coverage and regularization effects. To model the background, we added a constant intensity template. This template is typically required to be included in any analysis to ensure reliable feature extraction.

\subsubsection{Composite Templates}

A general image reconstruction from a VLBI observation may have multiple image features. As such, \VIDAjl allows the user to combine multiple features into composite templates, where each individual component is given a relative flux\footnote{Note the absolute or total flux of the image is not recoverable since we renormalize each image to have unit flux.}. An example of this is shown in \autoref{ssec:dbl} where three Gaussian templates are used to model the reconstructed image.

\subsection{Probability Divergences}\label{ssection:divergences}
As mentioned above, \VIDAjl uses an analogy between images and 2-D probability distributions to motivate the use of divergences as objective functions. Divergences form a general way to measure the similarity between two distributions. A divergence can be thought of as a functional $F_q[p] = \Dc(p||q)$, comparing the distribution $p$ (template) to a reference  $q$ (image), and is required to be non-negative, $\Dc(p||q)\geq 0$, and non-degenerate, $\Dc(p||q)=0$ if and only if $p=q$. Note that this definition is more general than a metric. Namely, a divergence does not have to be symmetric, i.e. $\Dc(p||q)\neq \Dc(q||p)$ or satisfy the triangle equality.

One of the most well-known divergences is the Kullback-Leiber (KL) divergence \citep{kullback1951} or relative entropy,
\begin{equation}
    KL(p||q) = \int p(\bm{x})\log\left(\frac{p(\bm{x})}{q(\bm{x})}\right)\dd^2 \bm{x}.
\end{equation}
One issue with the KL divergence is its definition when the support of q and p differ, i.e. if $q(x)=0$. In this case we set the contribution to the integral to be zero.

In addition to the KL divergence \VIDAjl also includes the Bhattacharyya divergence (Bh) \citep{Bhattacharyya},
\begin{equation}\label{eq:bhdiv}
    Bh(p||q) = -\log\int\sqrt{p(\bm{x})q(\bm{x})}\dd^2 \bm{x}.
\end{equation}

The Bh divergence is related to a well known metric on probability spaces, the \textit{Square Hellinger Distance
}:
\begin{equation}
    H(p,q) = \frac{1}{2}\int (\sqrt{p(\bm{x})} - \sqrt{q(\bm{x})})^2 \dd^2 \bm{x} = 1-Bh(p||q). 
\end{equation}
Therefore, minimizing the Bh divergence is simply a least squares fit in the space of the square root of distributions.

In this paper we will present the results from optimizing the Bh divergence for two reasons. First, we found that while the KL and Bh divergence produce near-identical results, the Bh divergence required $\sim 25\%$ less evaluations to converge. Second, the Bh divergence has preferable theoretical properties compared to the KL divergence. Namely, it is well defined when the image pixels have zero intensity and is symmetric.

\subsection{Optimizing the Divergence}

A problem when using probability divergences is that they give non-convex, non-linear optimization problems. Furthermore, the nature of the problem will change if the template changes, making an analytic analysis difficult. Therefore, to extract the globally optimal template, we turned to heuristic global optimizers, such as Genetic/Evolutionary strategies \citep[see ][ for a review]{diffevo} and simulated annealing \citep{samin}. For this paper we used the Julia package BlackBoxOptim.jl\footnote{https://github.com/robertfeldt/BlackBoxOptim.jl} \footnote{\VIDAjl also has interfaces to other Julia optimization packages, such as Optim.jl and CMAESEvolutionStategy.jl}. BlackBoxOptim.jl uses natural-evolution, and differential evolution strategies to perform a stochastic search of the parameter space. In the next section we will validate that our chosen optimizer is able to recover the optimal template reliably.

\begin{figure*}[!ht]
    \centering
    \includegraphics[width=\textwidth]{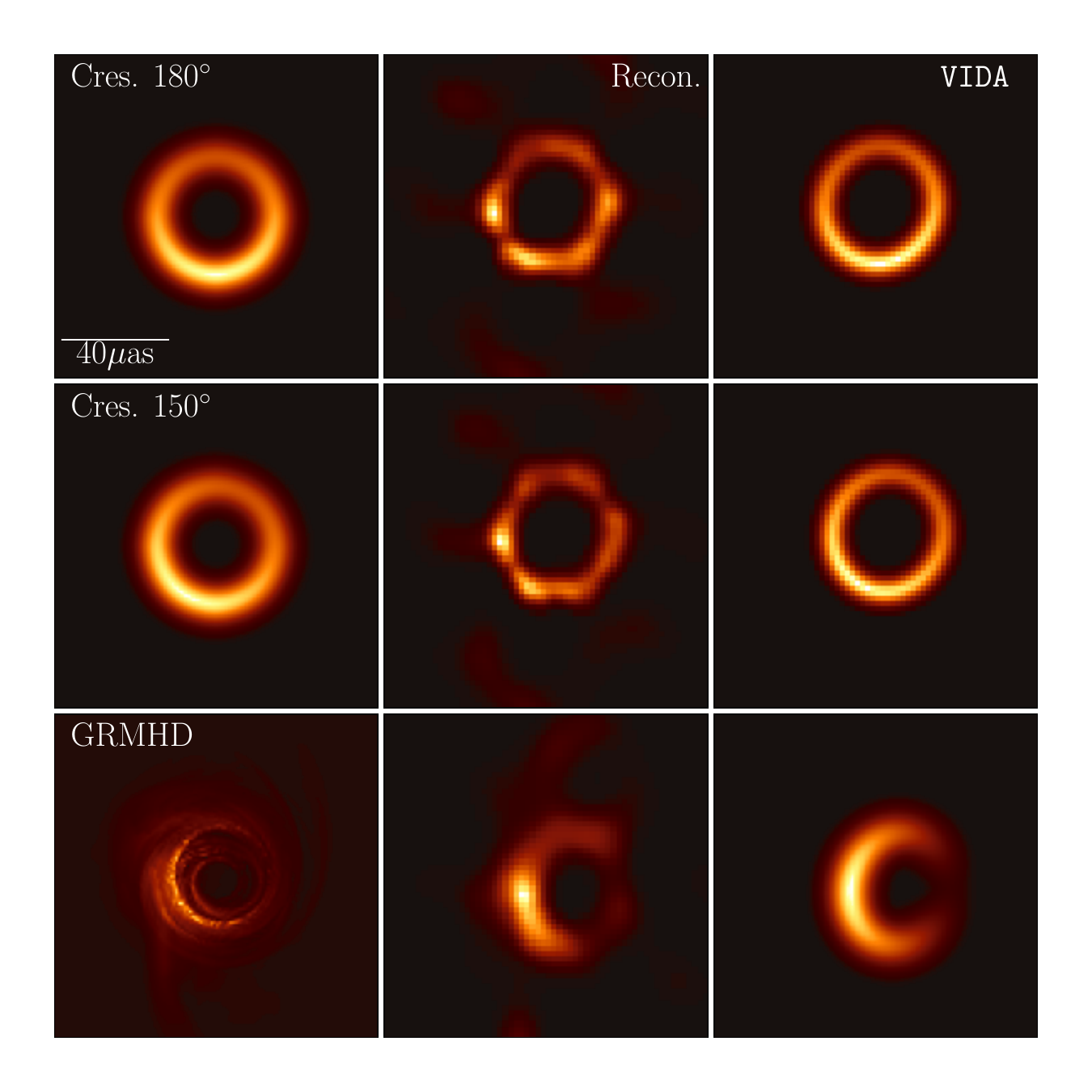}
    \caption{Images used for the imaging validation from \citetalias{EHTCIV}. We considered 3 models: \textit{top row} crescent with position angle $\xi=180^\circ$ north of east, \textit{middle row} crescent with position angle $\xi=150^\circ$ and    \textit{bottom row} GRMHD simulation.  The left column shows the truth image, the middle columns an example reconstruction, and right the optimal \VIDA template applied to the reconstruction.}\label{fig:geom_test}
\end{figure*}

\section{Validating \VIDA}\label{sec:validation}

To validate \VIDA we need to analyze two related quantities. First, we need to verify that the objective function, i.e. the Bh divergence \autoref{eq:bhdiv}, is robust to the artifacts that occur in image reconstructions. Namely whether the recovered parameter distribution contains the true value. Additionally, given the complex nature of the optimization problem we need to ensure that the chosen optimizer can recover the global minimum of the divergence. Our validation procedure will consist of:
\begin{enumerate}
    \item Selecting an applicable ground truth image $I_{\rm truth}$ (see the left column of \autoref{fig:geom_test} 
    \item For each truth image create a simulated EHT observation matching the observation characteristic of the EHT \m87 2017 observations.
    \item Create an ensemble of image reconstructions of the simulated observations using the same procedure as \citetalias{EHTCIV}.
    \item Apply VIDA to each image reconstruction and compare the inferred results to the ground-truth parameters, and the reconstruction technique \ReX used in \citetalias{EHTCIV} and \citetalias{EHTCVI}.
\end{enumerate}

\subsection{Step 1: Selecting Ground Truth Images}
To validate \VIDA, we applied it to a subset of the test set from \citet{EHTCIV}.
The sources we considered are shown in \autoref{fig:geom_test}, and consist of two geometric crescents and a general-relativistic-magneto-hydrodynamical (GRMHD) simulation from \citet{EHTCV}. The geometric crescent model is described by:
\begin{equation}
    I(r,\theta) = F_0(1 - s \cos(\theta - \xi))\frac{\delta(r-r_0)}{2\pi r_0},
\end{equation}
The infinitely thin ring is then convolved with a circular Gaussian with FWHM $10\,\muas$. Two orientations (measured east-of-north) $\xi=180^\circ$ and $\xi=150^\circ$ are considered in this paper, and are shown in the top and left panels of \autoref{fig:geom_test}. For both orientations we took $r_0 = 22\, \muas$, $s=0.46$, and \textbf{$F_0 = 0.6\,\rm Jy$}. After blurring the ring, it is important to note that the effective radius (the intensity peak) is smaller than the original radius of the ring \citepalias[see][]{EHTCIV}. The amount the diameter is biased inwards is given approximately by:
\begin{equation}\label{eq:diam_bias}
    d_{\rm blur} = d_{\rm true} - \frac{1}{8\ln 2}\frac{\alpha^2}{d_{\rm true}},
\end{equation}
where $d_{\rm true}$ is the diameter of the non-convolved ring (44\, \uas), and $\alpha$ is the FWHM of the Gaussian kernel ($\alpha$ = 10\,\uas). Using \eqref{eq:diam_bias} with $d_{\rm true}=44\,\muas$ gives $d_{\rm blur} \approx 43\,$\uas. If we also consider the finite resolution of the EHT array ($\sim 20\muas$) this is further decreased to $\approx 42\,$\uas. Therefore, we expect both \VIDA and \ReX to recover a diameter of $42\,\muas$. Additionally, the slash strength is also modified by the convolution. Fitting the crescent with the GGR template we find $s=0.32$, which is the value we will take as the ground truth below.

\begin{figure*}
    \centering
    \includegraphics[width=1.0\linewidth]{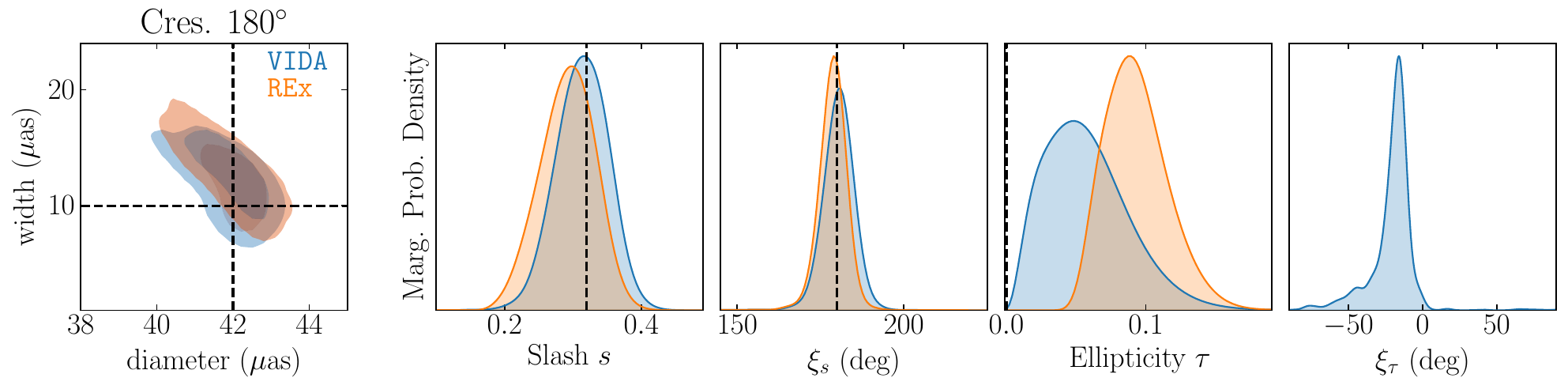}\\
    \includegraphics[width=1.0\linewidth]{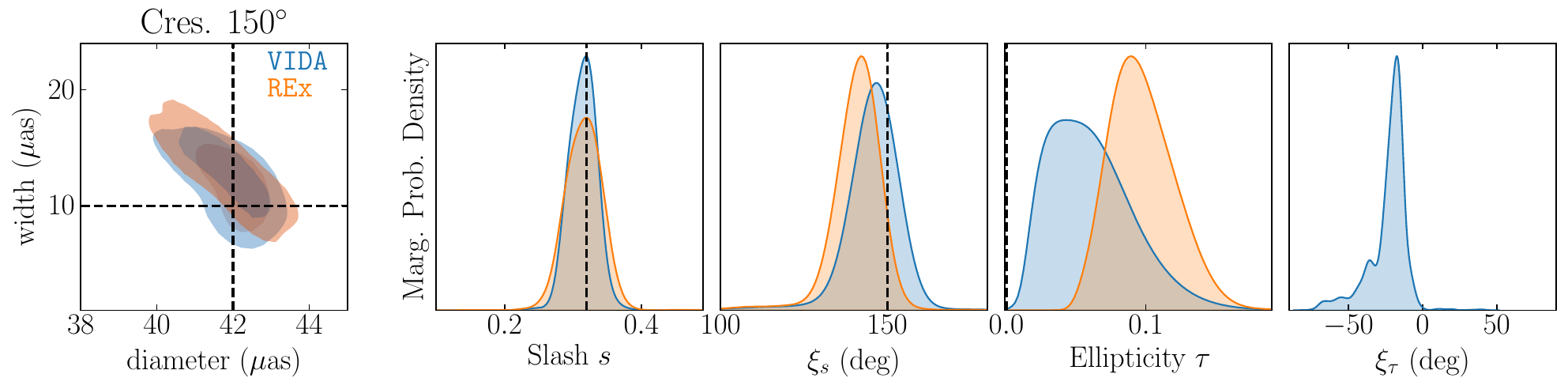}\\
    \includegraphics[width=1.0\linewidth]{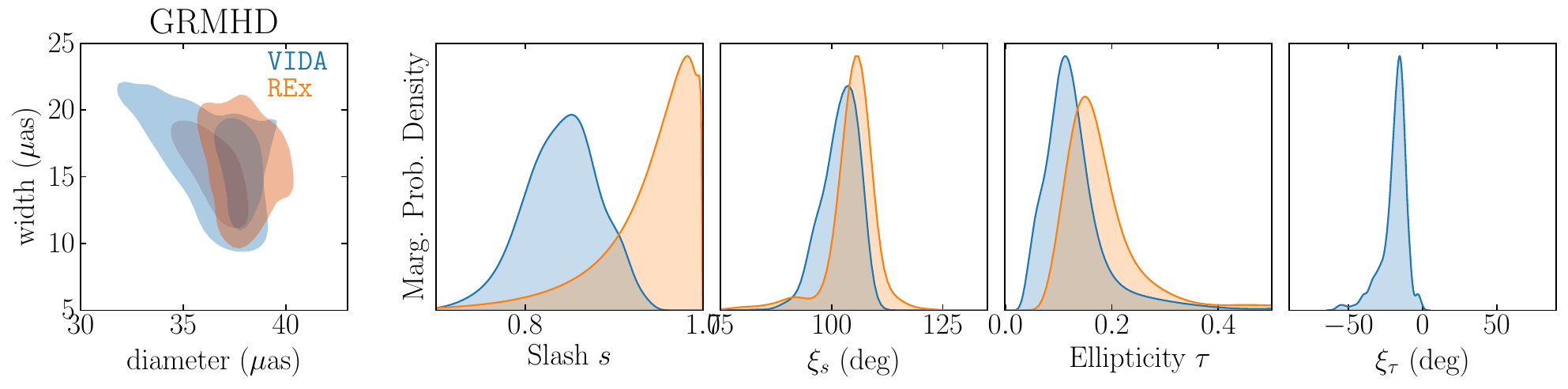}\\
    \caption{Results from \VIDA (blue) and \ReX (orange) being applied to the \citetalias{EHTCIV} reconstructions using the simulated April 11 EHT sampling of \autoref{fig:geom_test}. The top and middle rows show the geometric crescent results and the bottom the GRMHD. For the crescent models we found good agreement between \VIDA (blue) and \ReX (orange) for the ring diameter, width, slash, and location of the azimuthal brightness. Furthermore, the crescent distributions for the diameter (after accounting for \autoref{eq:diam_bias}), width, slash strength $s$ and position angle $\xi_s$ are consistent with the truth image (black dashed line). For the ellipticity, $\tau$, we found \VIDA's value was systematically lower than \ReX's as expected since \ReX measures the fractional dispersion of the ring radius and random fluctuations can then create a ellipticity floor around 0.05 (see \autoref{appendix:rex} for a detailed discussion). The ellipticity orientation, $\xi_\tau$, is only recovered by \VIDA, so there is no \ReX comparison. For the GRMHD simulations, the agreement between \ReX and \VIDA is weaker. Namely, the distributions for the parameters are sometimes significantly different, although the portions of the distributions do overlap. Similar results were found using the simulated EHT coverage from April 5, 6, and 10.}
    \label{fig:m87geom_results}
\end{figure*}

\subsection{Step 2: Creating Simulated EHT Observations}
While \VIDA could be applied to the ground-truth images shown in the left column of \autoref{fig:geom_test}, this isn't applicable to what the EHT observes. The EHT is a very-long-baseline inferometer and instead observes complex visibilties, $V(u,v)$ which are related to the on sky image through the van Cittert–Zernike theorem \citep{TMS}:
\begin{equation}
    V(u,v) = \int e^{-2\pi i (u\alpha + v\beta)}I(\alpha,\beta)\dd \alpha \dd \beta.
\end{equation}
In addition, atmospheric and telescope effects can further corrupt the signal. To model these corruption effects we use the \ehtim package \citep{Chael_2016, Chael_2018} to generate realistic simulated data.

\subsection{Step 3: Generating an Ensemble of Reconstructed Images from Simulated VLBI Data}\label{ssection:topset}
 To validate VIDA we used image reconstructions from the forward modeling or ``regularized maximum likelihood methods'' (RML), e.g. \citet{Honma2014, Bouman_2016, Akiyama_2017a,Akiyama_2017b, Ikeda2018, Kuramochi_2018} and more specifically the \texttt{eht-imaging} package \citep{Chael_2016,Chael_2018}.  The goal of RML methods is to find the image, $I$, that minimizes the objective function
 \begin{equation}\label{eq:img-objective}
     J(I) = \sum_{\rm data} \alpha_d \chi^2_d(I) - \sum_{\rm regularizers}\beta_r S_r(I).
 \end{equation}
 Following \citetalias{EHTCIV}, each $\chi_d^2$ is defined solely from the data products from the EHT telescope, e.g., complex visibilities. The second term encapsulates our additional assumptions, or regularizers, that are placed on the image. The $\alpha_d\, \beta_r$, are the ``hyperparameters'' that control the relative weighting of the regularizers and data products. For the list of the regularizers used, see \citetalias{EHTCIV}. In an attempt to model the uncertainty in the image reconstructions, we used the same set of imaging hyperparameters as \citetalias{EHTCIV}. The resulting set of image reconstructions is called the ``topset'' and results in 1572 reconstructions per dataset.
 
 For completeness we also apply \VIDA to the CLEAN \m87 reconstructions for the Crescent 180, GRMHD, double and Disk in \autoref{appendix:clean}. Note that for \m87 there are only 30 reconstructions for each simulated dataset, making the comparison more uncertain.

\begin{figure*}[!ht]
    \centering
    \includegraphics[width=\linewidth]{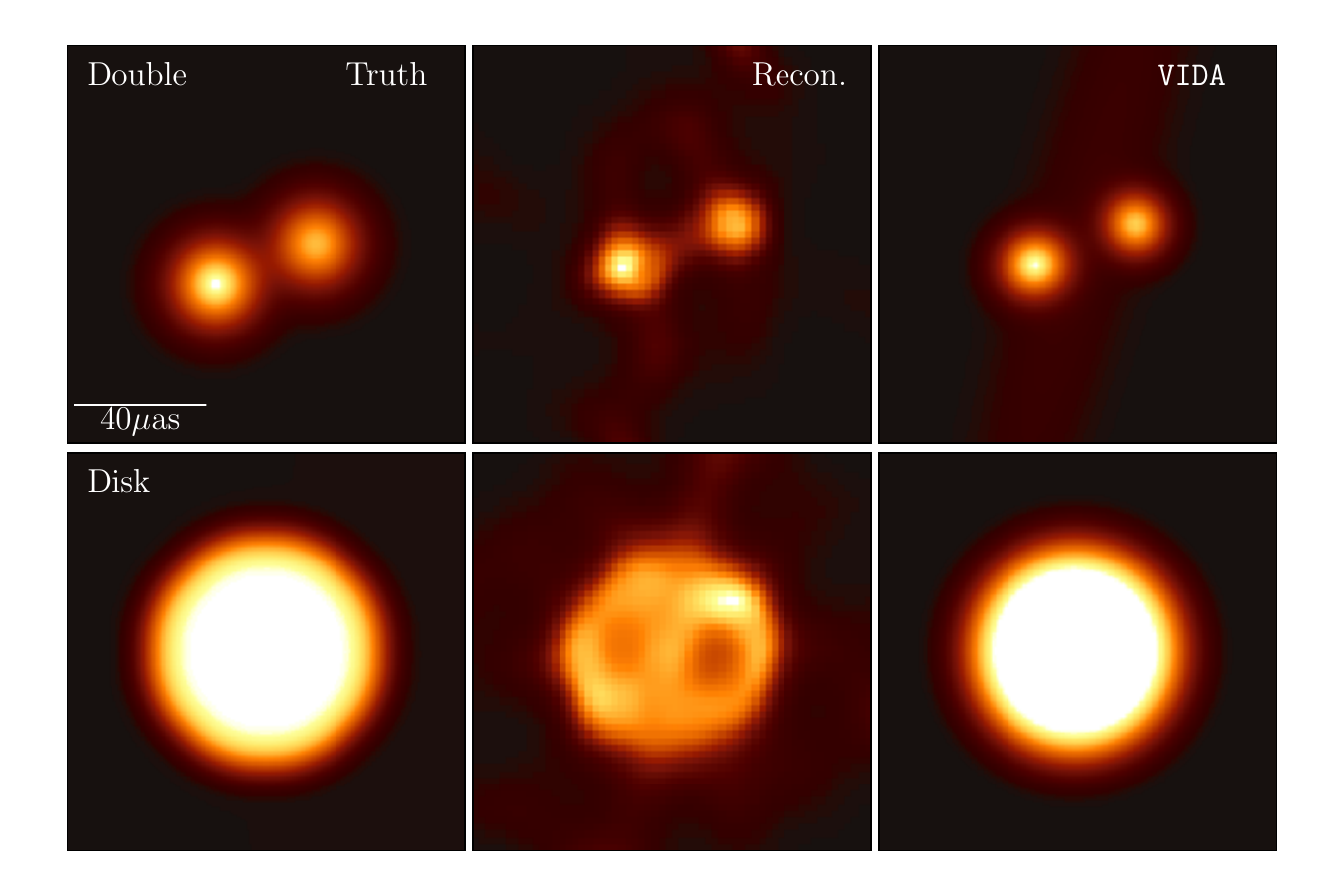}\vspace{-4mm}
    \caption{Results from applying \VIDA to the non-ring test images from \citetalias{EHTCIV} where the left column shows the ground truth image, the middle an example reconstruction from the topset, and right the optimal template from applying VIDA to the reconstruction. \textit{Top}: Shows the results for the double image. \textit{Bottom}: Shows the results for the disk. Overall \VIDA can recover the intrinsic structure of both images.}
    \label{fig:non_ring_res}
\end{figure*}

\subsection{Step 4: Applying \VIDA to the Ring-Like Image Reconstruction Ensembles}\label{ssection:paperiv}

\VIDA was run on each set of image reconstruction ensembles using the GGR template with a constant brightness background whose relative flux was a free parameter, giving 9 parameters in total. Some example reconstructions and corresponding optimal templates are shown in \autoref{fig:geom_test}. The results are shown in \autoref{fig:m87geom_results}. For the crescent models, we were able to recover the expected diameter $d$, width $w$, and azimuthal orientation $\xi_s$ (black dotted lines in \autoref{fig:m87geom_results}. In addition, we compared the \VIDA results to the \ReX method used in \citetalias{EHTCVI}.

\ReX assumes that a single ring-like feature dominates the image reconstruction, and then finds the ring by finding the image location that leads to a ring with minimal radial dispersion. \ReX characterizes (see \autoref{appendix:rex} and \citetalias{EHTCIV} for definitions) the ring through a diameter $d$, width $w$, brightness moments $s$ and orientation $\xi_s$ and a fractional dispersion of the diameter $f_d$. The diameter and width and brightness profile of the \ReX measurement are similar to \VIDA's measurement with the GGR template. However, the fractional dispersion is not directly measured. Instead, \VIDA measures the ellipticity of the ring. In \autoref{appendix:rex} we demonstrate how $f_d$ and $\tau$ are related if the dominant source of radial dispersion in the ring is due to ellipticity. 

 The agreement between \ReX and \VIDA is excellent for both the crescent images. The peak and overall width of the distribution for each parameter in \autoref{fig:m87geom_results} are consistent between \ReX and \VIDA. The ground truth values (black vertical lines) for the diameter, width, and brightness orientation $\xi_s$ are also consistent with the \ReX and \VIDA results. We also analyzed the pairwise linear-correlations between all the parameters and found that only the diameter and width were correlated (see \autoref{fig:paircorr}). This correlation is expected from \autoref{eq:diam_bias} and occurs due to the finite resolution of the EHT array.

\begin{figure*}[!ht]
    \centering
    \includegraphics[width=\textwidth]{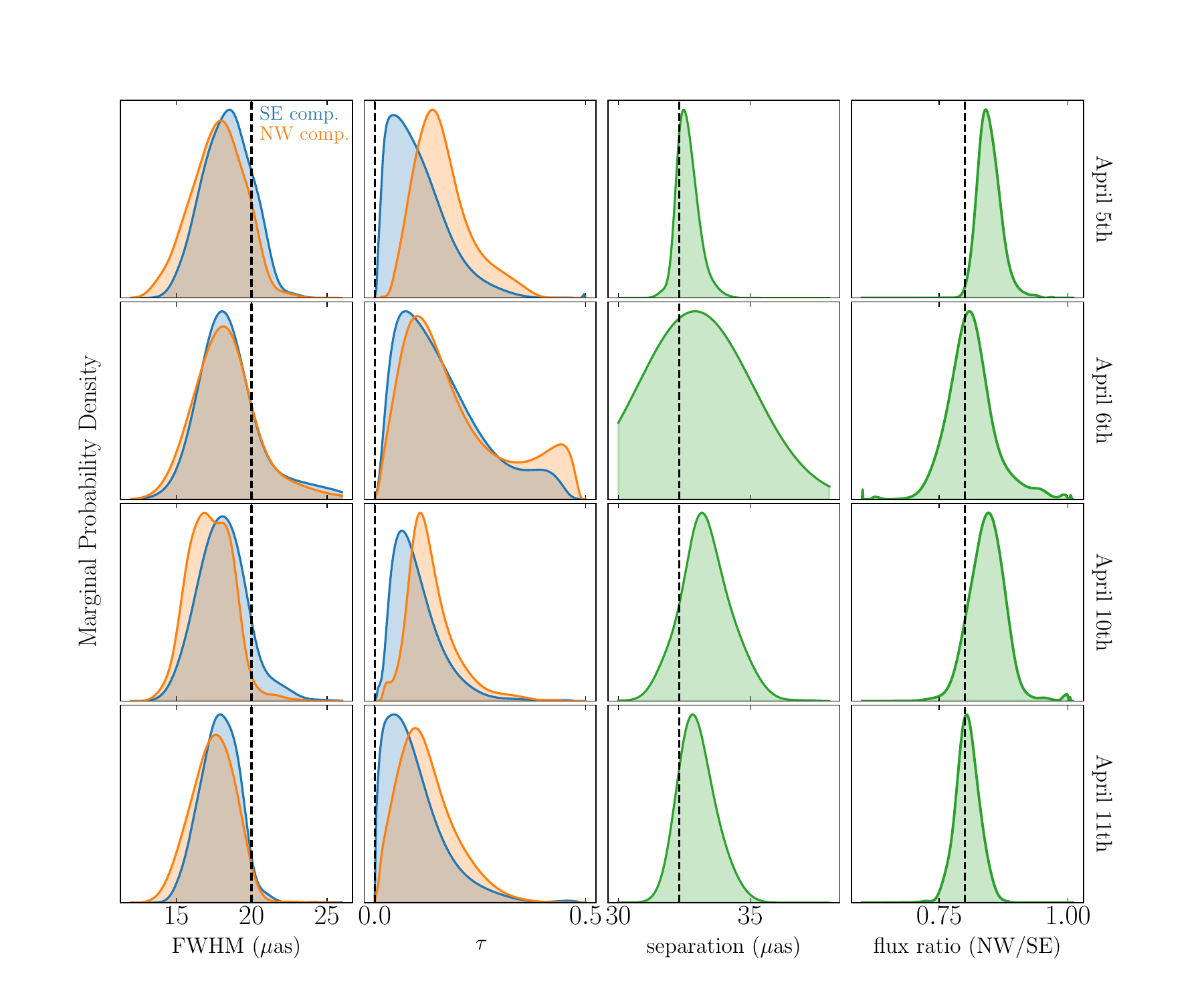}
    \caption{\VIDA results for the two compact Gaussian components (blue for SE component and orange for the NW) in the double Gaussian test image. Each row corresponds to a different EHT sampling from the 2017 \m87 observations. The green curves are for parameters that are a combination of the SE and NW components. On all days the true values are included in the support of the parameter distributions found by \VIDA. Note, that the broad distribution found using the simulated EHT April 6 sampling is due to an imaging artifact as discussed in the paper.}
    \label{fig:dbl_marg}
\end{figure*}

To compare \VIDA and \ReX's measurement of ellipticity, we first note that an additional processing step is needed since the two definitions differ. \ReX doesn't directly measure $\tau$ but instead measures the fractional diameter dispersion of the ring $f_d$ (see \autoref{eq:rex_asym} for a definition). If we assume that the ring's ellipticity dominates $ f_d$, then $f_d$ and $\tau$ are related by an invertible map. For more information about this conversion, see \autoref{appendix:rex}. In \autoref{fig:m87geom_results} we show \ReX's results after converting from $f_d$ to $\tau$. Comparing the two measurements of $\tau$, we see that \ReX's measurement is consistently greater than \VIDA's. This bias is not unexpected given that when $\tau$ is small, the conversion described in \autoref{appendix:rex} no longer applies. Instead, the fractional dispersion is dominated by random fluctuations in the ring diameter, creating a floor in $f_d$. If we then naively apply the previous conversion, as was done in \autoref{fig:m87geom_results}, we will overestimate $\tau$ (see \autoref{fig:asym_cal}). 

\VIDA also recovers the orientation of the ring ellipticity $\xi_\tau$. Interestingly, in all instances, we measure a similar distribution for $\xi_\tau$ irrespective of the intrinsic image.
This distribution is biased so that the semi-major axis of the ellipticity is in the north-south direction. By visually inspecting the image reconstruction ensembles (see \autoref{fig:geom_test} for a typical example reconstruction) we confirmed this feature was present in the reconstructions and was not a bias from \VIDA. We also find a non-linear correlation between the measured ring ellipticity and orientation. Namely, $\tau$ is significantly larger when $\xi_\tau \approx 0$. The origin of this bias and implications for \m87 will be discussed in \citet{VIDAII}.

\section{Applying \VIDA to Additional Simulated Image Reconstructions}\label{section:asym}
In the previous section, we saw that \VIDA and \ReX gave remarkably similar answers to problems that demonstrated similar ring-like structures. While \ReX is limited to ring extraction, \VIDA can be applied to any image given a suitable template function. This section will explore \VIDA's capabilities of extracting features from a broader range of potential sources. To accomplish this, we will consider the other non-ring test images from \citetalias{EHTCIV}: the symmetric disk and double Gaussian (see \autoref{fig:non_ring_res}). We will follow the same steps in the previous section to evaluate \VIDA's performance.

\begin{figure*}[!ht]
    \centering
    \includegraphics[width=\linewidth]{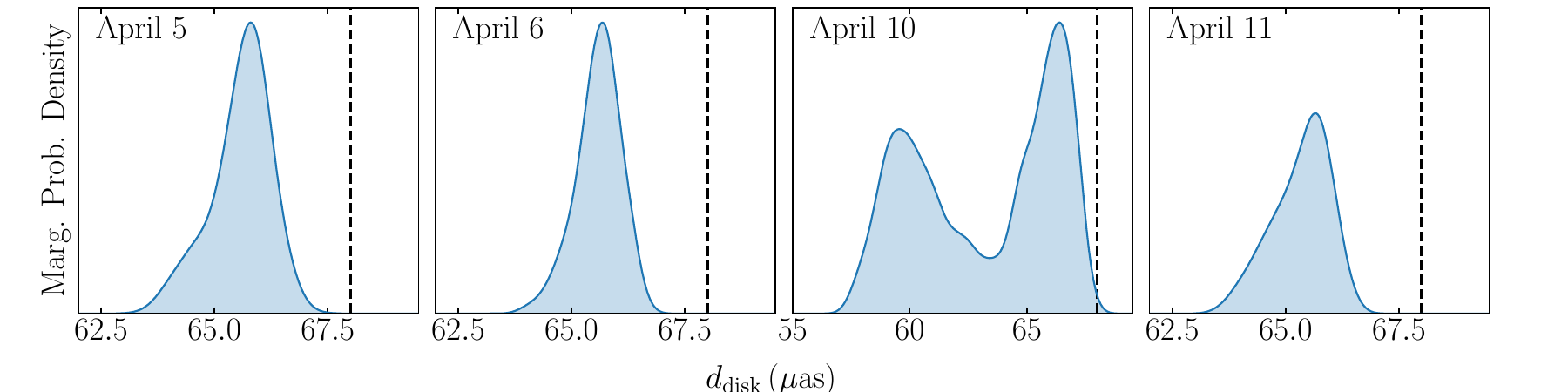}\vspace{-2mm}
    \caption{\VIDA results of the diameter for the disk topset. The diameter is given by \eqref{eq:disk_diam}. Ignoring the simulated EHT April 10 sampling, we consistently find that the diameter is $4\muas$ smaller than the original image. The origin of this discrepancy is discussed in \autoref{fig:avg_disk_trip}. For the simulated data using the April 10 sampling, which has poor coverage compared to the other days, the imaging gives two modes. One mode is similar to the other days, while the second fails to show a coherent disk structure giving the second peak in diameter at $60\muas$.}
    \label{fig:disk_marg}
\end{figure*}

\subsection{Double Gaussian}\label{ssec:dbl}
Here we consider a source composed of a compact double with two circular Gaussian components. Each Gaussian has an FWHM of $20\,\muas$. One of the Gaussians is placed at the origin and has a flux of $0.27\rm Jy$, which we will call the NW component. The other Gaussian is at $\Delta\rm{RA} = 30\muas$ and $\Delta\rm{DEC} = -12\muas$ and has a flux of $0.33\rm Jy$ and will be called the SE component. This type of source could arise when looking at AGN using VLBI images, such as the recent 3C~279 results \citep{3c279EHT}. 

To extract the reconstruction's compact components, we used a template with three asymmetric Gaussian components and constant background. Two of the Gaussian components were allowed to be arbitrary, while the third Gaussian component was forced to be large ($r_0 > 15\,\muas$). The reason for the third Gaussian was there tended to be a region of additional emission around the two dominant Gaussian components in the image reconstructions. This diffuse emission can be seen in the top middle panel of \autoref{fig:non_ring_res}. If we didn't include this third component, we found that the Gaussian components tended to be quite large to soak up the extra emission.

\VIDA's results for the double Gaussian are shown in \autoref{fig:dbl_marg}. Overall the on-sky size of each Gaussian, their separation, and flux ratio are contained in the support of the distribution. However, there do exist biases in the results, such as the FWHM of the Gaussian being biased low\footnote{Note that the FWHM of the Gaussian's is related to the hyperparameters of the top set.}. Additionally, the ellipticity $\tau$, is larger than zero; however, this is to be expected as asymmetry can only be added to the images.

For the simulated EHT sampling from April 6, we see that the ellipticity appears to be bimodal, and the parameter uncertainties are greater than the other days. This uncertainty was unexpected given the relatively good EHT coverage on April 6. After analyzing the image reconstructions we deemed this not to be due to \VIDA. Instead, a subset ($10-15\%$) of image reconstructions on April 6 that exhibit a third bright Gaussian feature. If we remove these reconstructions, we find that the results for the simulated April 6 coverage are consistent with the other days.

\subsection{Disk Image}\label{ssec:disk}
The intrinsic image is a symmetric flat disk with a diameter of $70\,\muas$, which is then convolved with a Gaussian with an FWHM of $10\,\muas$. The true image, an example reconstruction, and optimal template for that reconstruction are shown in the lower left, middle panel, and right panel of \autoref{fig:non_ring_res}.

To encode the diameter of the disk we use FWHM of the disk template:
\begin{equation}\label{eq:disk_diam}
    d_{\rm disk} = 2r_0 + 2\sqrt{2\log2}\alpha,
\end{equation}
where $r_0$ and $\alpha$ are described in \autoref{eq:disk_template}. We fit the disk template to the ground truth image to calibrate the diameter definition to the disk's true diameter. We found that the optimal template for the true image had a $d_{\rm disk} \approx 69\,\muas$. If we convolved the image by an additional $20\,\muas$ to take into account the finite resolution of the EHT array, we found $d_{\rm disk} \approx 68\,\muas$. This is the value we use as the ground truth diameter in all comparisons below.

\autoref{fig:disk_marg} displays the results for \VIDA applied to each day. Ignoring the simulated EHT sampling of April 10, which has poor coverage compared to the other days, we find that the results are very consistent between days. For the simulated April 5 EHT sampling, we find the median diameter $d_{\rm disk} = 65.7^{+0.42}_{-0.76}\,\muas$ where the range are the $68\%$ interval about the median. Similarly, for the simulated EHT sampling of April 6th, we find $d_{\rm disk} = 65.6^{+0.40}_{-0.47}\,\muas$, and for April 11th $d_{\rm disk} = 65.5^{+0.45}_{-0.80}\,\muas$. This demonstrates that \VIDA is robust to the slight difference in image reconstructions from different baseline coverage. 

For the simulated EHT sampling of April 10, we had a different result finding a bimodal diameter. Analyzing the reason for this, we found that images with $d_{\rm disk} \approx 60\,\muas$ had a markedly different structure than the rest of the images. Given the distinct non-disk structure of the image, it is no surprise that \VIDA struggles at recovering the correct diameter.

\begin{figure}[!ht]
    \centering
    \includegraphics[width=\linewidth]{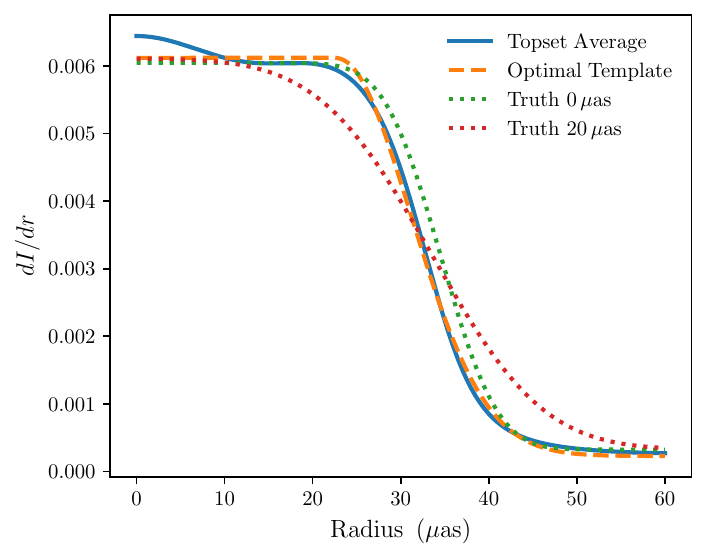}
    \caption{Results when applying \VIDA to the average reconstruction from the topset using the  simulated April 11 EHT sampling. The images were normalized to unit flux and centered before averaging. The average radial profile is shown in solid blue. Comparing this to the optimal template (orange dashed line) and the true profile (dotted lines), we see that the optimal template matches the reconstruction's radial profile but underestimates the ground truth image.}
    \label{fig:avg_disk_trip}
\end{figure}

Comparing our result for the diameter to the true value, $68\muas$, we find that our result has a consistent bias of $\approx 2.4\muas$ for the simulated April 5, 6, and 11 EHT sampling. Again, this appears to be an artifact of the imaging process. In \autoref{fig:avg_disk_trip}, the radial profiles of the truth (dotted lines), averaged image reconstructions\footnote{When averaging we first centered the images by computing the image centroid and normalized the images to have unit flux.}, and optimal templates are compared.
\VIDA does an excellent job of recovering the size of the images, which are similarly biased toward smaller radii. This suggests that the diameter bias is intrinsic to the topset used for this disk. Furthermore, as shown in \autoref{appendix:clean} the CLEAN reconstructions of the disk do not suffer from the same bias. Given this, it is unlikely that observed bias is due to the finite resolution of the EHT array and is intrinsic to the top set used.

\section{Summary and Conclusions}\label{sec:conclusions}

We present \VIDA, a new image feature extraction technique appropriate for use by the EHT.  \VIDA adopts a forward modeling approach to extract quantitative image properties by approximating the image with a parameterized family of functions that encode the desired image properties.

A key feature of \VIDA is its flexibility. Multiple image components have already been implemented, from which composite models of significant complexity can be constructed.  These include ring-like templates of particular relevance to EHT images, Gaussians, and constants.

The ability of \VIDA has been demonstrated for several sources, each with over a thousand reconstructions.  These include image reconstructions from simulated data produced from double Gaussians, slashed rings, and GRMHD simulations.  In all cases, key quantitative features were accurately recovered where they appeared in the underlying image reconstructions. These include separations, orientations, ring diameters, widths, brightness profiles, and multiple measures of ellipticity.  Application of these to the EHT observations of \m87 will be explored in future work.

The applicability of \VIDA extends beyond EHT observations of \m87.  The ability to create composite models with multiple components is naturally relevant to VLBI reconstruction of AGN, such as 3C~279, that is composed of multiple compact features \citep[e.g.,][]{3c279EHT}.

It should be noted that image feature extraction methods, like \VIDA, are generally most useful when strong priors may be placed on the image structure itself.  That is, \VIDA is primarily a method for quantifying what is already qualitatively apparent.  Poorly chosen models can lead to significant parameter biases, as seen in \autoref{ssec:dbl}, where an extra Gaussian blob was required to achieve acceptable results.  However, because \VIDA is an image characterization tool, not an imaging tool in and of itself, this presents only a very modest limitation on its utility.

\software{BlackBoxOptim.jl, \texttt{eht-imaging}, \texttt{GR} \citep{gr}, Julia \citep{julia}, matplotlib 3.3 \citep{matplotlib}, Pandas \citep{pandas, reback2020pandas}, Python 3.8.3 \citep{python3}, Scipy \citep{SciPy}, ThemisPy, VIDA.jl}

\acknowledgments
We thank George Wong, Ben Prather, and Charles Gammie for kindly providing synthetic images based on GRMHD simulations. We would also like to thank the referees for their helpful comments. This work was supported in part by Perimeter Institute for Theoretical Physics.  Research at Perimeter Institute is supported by the Government of Canada through the Department of Innovation, Science and Economic Development Canada and by the Province of Ontario through the Ministry of Economic Development, Job Creation and Trade.
PT receives support from the Natural Science and Engineering Research Council through the Alexander Graham Bell CGS-D scholarship.
A.E.B. thanks the Delaney Family for their generous financial support via the Delaney Family John A. Wheeler Chair at Perimeter Institute.
A.E.B. receives additional financial support from the Natural Sciences and Engineering Research Council of Canada through a Discovery Grant. We thank the National Science Foundation (AST-1716536, AST-1440254, AST-1935980) and the Gordon and Betty Moore Foundation (GBMF-5278) for financial support of this work. This work was supported in part by the Black Hole Initiative, which is funded by grants from the John Templeton Foundation and the Gordon and Betty Moore Foundation to Harvard University.

\bibliographystyle{aasjournal}
\bibliography{main}

\appendix
\section{\ReX ring parameter definitions}\label{appendix:rex}
 \begin{figure}[ht]
     \centering
     \includegraphics[width=0.5\linewidth]{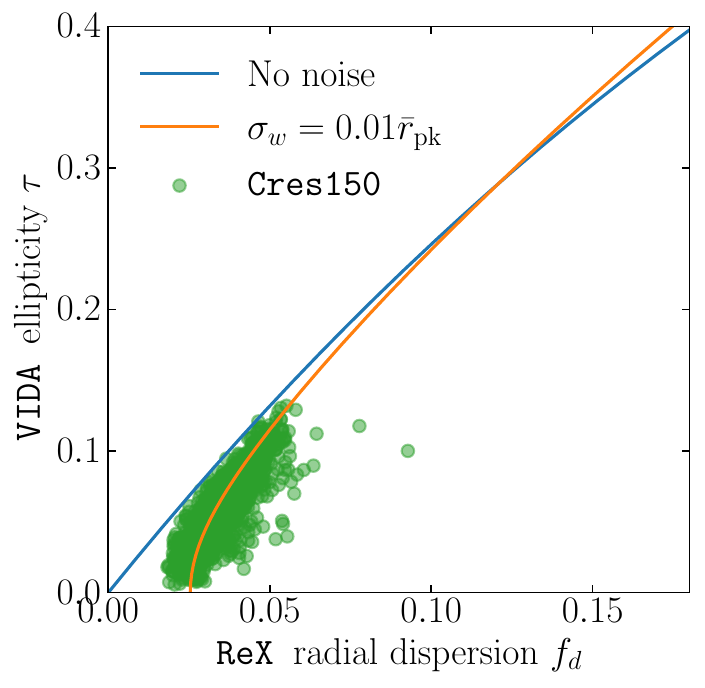}
     \caption{Comparison of the \ReX's fractional radial dispersion and \VIDA's $\tau$ ellipticity parameter. The blue curve shows the conversion for the case of an ellipse, the orange curve is an ellipse with a constant white noise fluctuation in the found radius, with a variance of $0.01^2\bar{r}_{\rm pk}^2$, and the green points are the results of fitting the Cres150 topset in \autoref{sec:validation} with both \ReX and \VIDA.}
     \label{fig:asym_cal}
 \end{figure}
The first step in \ReX \citepalias[see][for details]{EHTCIV} is to identify the dominant ring in the image. Given a center position $(x,y)$, \ReX samples the image in radius $r$ and azimuthal angle $\theta$ obtaining a intensity map $I(r,\theta|x,y)$. Then for that central map, the radius of the ring is taken as the azimuthally averaged peak brightness:
\begin{equation}
    \begin{aligned}
        r_{\rm pk}(\theta|x,y) &= \mathrm{argmax}[I(r,\theta|x,y)]_r\\
        \bar{r}_{\rm pk} &= \left<r_{\rm pk}(\theta| x,y) \right>_{\theta \in [0,2\pi]}.
    \end{aligned}
\end{equation}
This provides a different ``radius'' for every point $(x,y)$ in the image. To find the dominant ring in the image the fractional radius spread is minimized:
\begin{equation}
    (x_0,\; y_0) = \mathrm{argmin}\left[\frac{\sigma_{\bar{r}(x,y)}}{\bar{r}_{\rm pk}(x,y)}\right],
\end{equation}
where $\sigma_{\bar{r}}(x,y) = \left<(r_{\rm pk}(\theta|x,y) - \bar{r}_{\rm pk})^2\right>$, is the radial dispersion. The diameter in the image is then given by:
\begin{equation}
    d = 2\bar{r}_{\rm pk}(x_0,y_0).
\end{equation}
To relate this to \VIDA's definition we consider an ellipse with semi-major axis $a$ and semi-minor axis $b$. Then \VIDA parameterizes this ellipse with $d_0 = 2r_0 = 2\sqrt{ab}$ and $\tau = 1-b/a$. The relationship between $r_0$ and $r_{\rm pk}$ is then given by
\begin{equation}\label{eq:rex_rad_conv}
    \bar{r}_{\rm{pk}} = \frac{r_0}{\sqrt{1-\tau}}\frac{1}{2\pi}\int_0^{2\pi}\sqrt{1 - \epsilon^2(\tau)\sin^2(\theta)} \dd \theta = \frac{2}{\pi}\frac{r_0}{\sqrt{1-\tau}}E\big(\epsilon(\tau)\big),
\end{equation}
where $E(x)$ is the complete Elliptic integral of the second kind and $\epsilon(\tau) = \sqrt{1-(1-\tau)^2}$ is the orbital eccentricity.

\ReX's measure of circularity of the ring is provided by the radial fractional dispersion:
\begin{equation}\label{eq:rex_asym}
    f_d = \frac{\sigma_{\bar{r}}}{\bar{r}_{\rm pk}}.
\end{equation}
To compare \ReX's ellipticity measure, $f_d$, to \VIDA's, we need to relate $f_d$ to $\tau = 1-b/a$ 
\begin{equation}\label{eq:rex_asym_conv}
    f_d(\tau) = \frac{\sqrt{1-\epsilon(\tau)^2 - 4/\pi^2 E^2(\epsilon)}}{\sqrt{1-\tau}} = \frac{\sqrt{(1-\tau)^2 - 4/\pi^2 E^2(\epsilon)}}{\sqrt{1-\tau}}
\end{equation}
\autoref{eq:rex_asym} is then used to convert this to a fractional diameter spread. Using linear interpolation we invert the function achieving a map from $f_d$ to $\tau$. One important thing to note is that this conversion assumes that the image is a perfect ellipse. In general, this will not be true for the image reconstructions and expect that the \ReX ellipticity may be different than the \VIDA measurement. To model this we consider $r_{\rm pk}$ modified by a white noise $\epsilon_\theta$ term with dispersion proportional to the average radius $\left<\epsilon_\theta\epsilon_{\theta'}\right> = \sigma^2_{w}\bar{r}^2_{\rm pk}\delta(\theta - \theta')$. In this base the average peak radius is unchanged since $\epsilon$ has mean $0$. However, the additional noise does impact the radial dispersion:
\begin{equation}
    \sigma_{\bar{r}} \to \sigma_{\bar{r}} + 2\left<\epsilon_\theta r_{\rm pk}(\theta)\right> + \left<\epsilon^2_{\theta}\right>.
\end{equation}
When we have a circular ring then this just becomes $\left<\epsilon^2_\theta\right> = \sigma_\epsilon^2\bar{r}_{\rm pk}^2$ adding a constant floor. \autoref{fig:asym_cal} shows the conversion when the ring is elliptical and compares it to the results of the Cres150 topset of \autoref{sec:validation}.

The width of the ring is defined by finding the FWHM at a fixed $\theta$ ray, and then averaging over $\theta$,
\begin{equation}
    w = \left<\mathrm{FWHM}_r[I(r,\theta|x_0,y_0) - I_{\rm floor}]\right>_{\theta}.
\end{equation}
The intensity floor is given by $I_{\rm floor} = \left<I(r=50\muas, \theta)\right>_\theta$ and is included to avoid biasing the measurement due to the low level emission present in the image. This is similar to including the constant intensity template during the \VIDA extraction.

In order to characterize the azimuthal profile of the ring ($\xi_s$ and $s$ for \VIDA) we consider the azimuthal moments of the ring. Namely, the orientation $\xi_s$ is given by:
\begin{equation}
    \xi_s = \left<\mathrm{Arg}\left[
            \int_0^{2\pi}I(r,\theta|x_0,y_0)e^{i\theta}\mathrm{d}\theta
            \right]       
            \right>_{r\in [r_{\rm in},r_{\rm out}]},
\end{equation}  
where $r_{\rm in} = (d-w)/2$ and $r_{\rm out} = (d+w)/2$. The strength of the slash is given by
\begin{equation}
    s = 2\left<\frac{\left|\int^{2\pi}_{0}I(r,\theta|x_0,y_0)e^{i\theta}\mathrm{d}\theta \right|}{\int^{2\pi}_{0}I(r,\theta|x_0,y_0)\mathrm{d}\theta}\right>.
\end{equation}
Note that the factor of $2$ is included to match \VIDA's definition (\autoref{eq:slash}). 

\clearpage
\section{Joint Density Plot for Crescent 150 Reconstruction}

\begin{figure*}[!h]
    \centering
    \includegraphics[width=\linewidth]{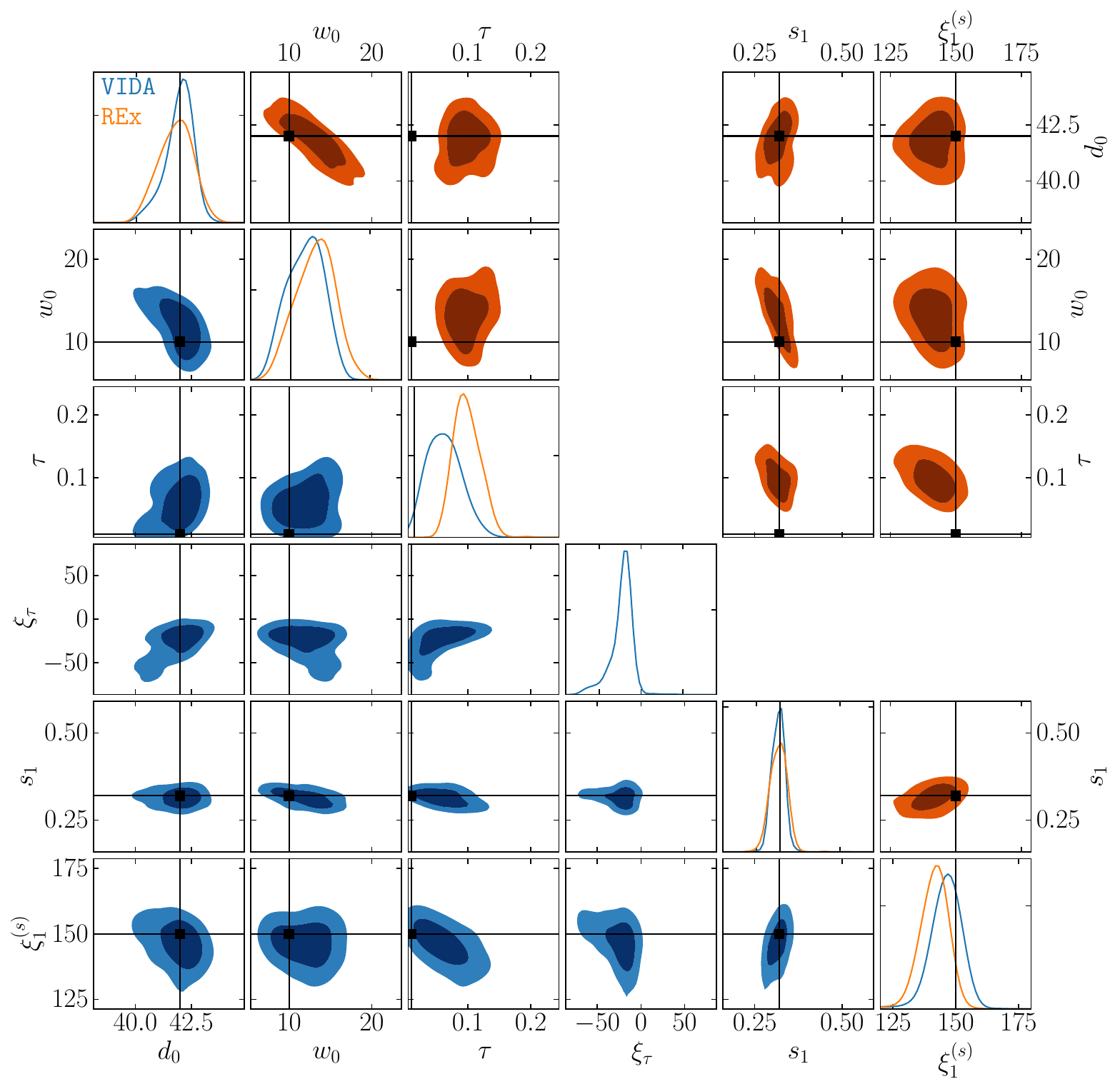}
    \caption{Joint density plot for the crescent image reconstruction ensemble using the April 11 EHT simulated coverage. For \VIDA (blue, lower), we find that the truth values (black lines) are within the recovered distributions. For \ReX (orange, upper) we find that for all parameter except the ellipticity $\tau$ are recovered. \ReX's ellipticity has a floor around $\tau=0.05$ which is expected since \ReX measures fractional dispersion as is explained in \autoref{appendix:rex}. Additionally, no $\xi_\tau$ is measured for \ReX so those columns are not shown.}
    \label{fig:paircorr}
\end{figure*}

\clearpage
\section{CLEAN Reconstruction Results}\label{appendix:clean}
\begin{figure*}[!t]
    \centering
    \includegraphics[width=\linewidth]{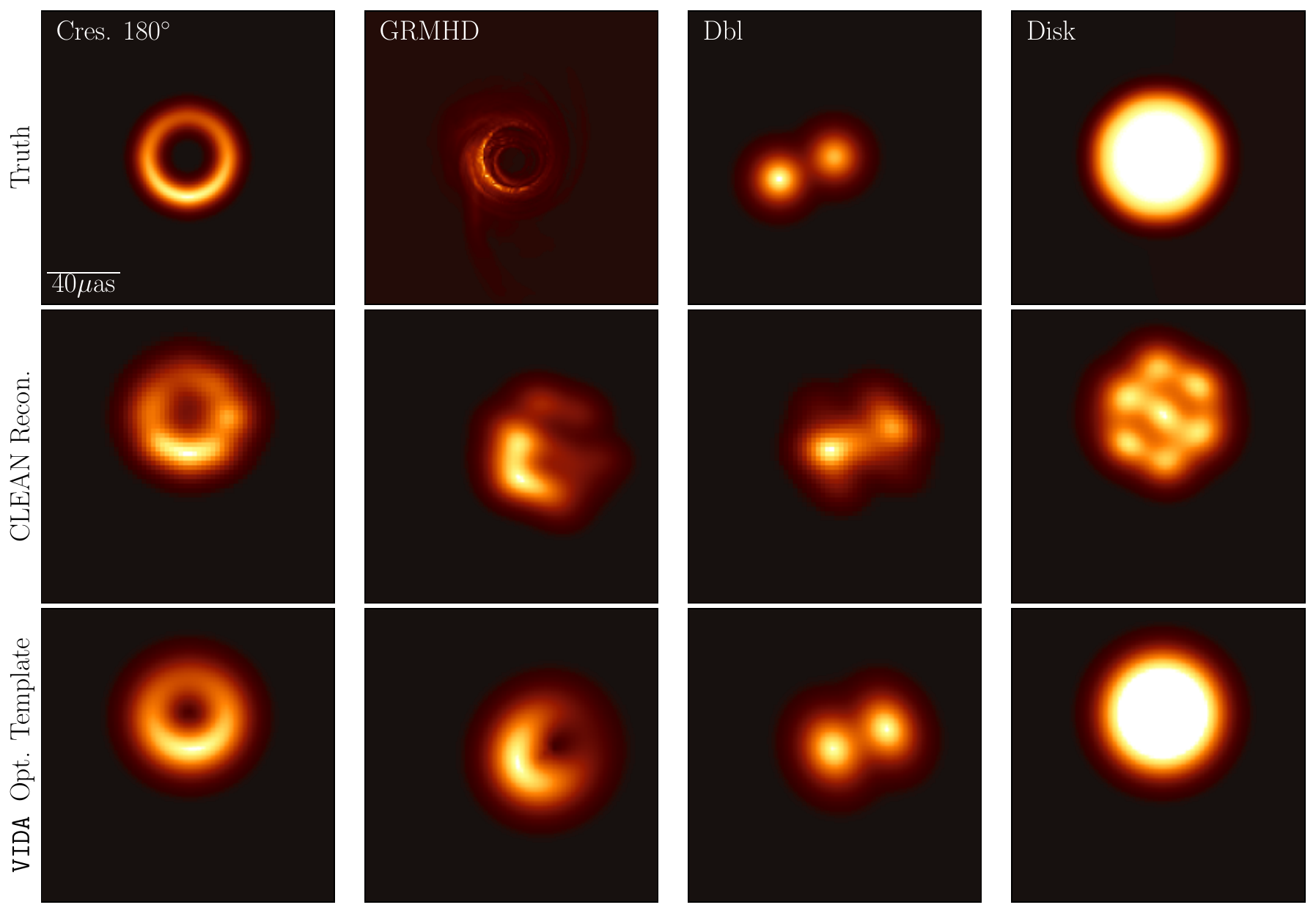}
    \caption{Example reconstructions and optimal \VIDA templates when applied to the crescent $180^\circ$ (first column), GRMHD simulations (second column), double (third column), and disk (fourth column). The top row shows the on-sky image, the middle row an example CLEAN reconstruction, and the bottom row the optimal \VIDA template.}
    \label{fig:clean_recon}
\end{figure*}

In this paper we have focused on RML methods. The reason for this twofold. First, the RML results consisted of the majority of the methods used in \citetalias{EHTCIV}. Second, and more important, RML methods produced thousands of simulated image reconstructions. The DIFMAP/CLEAN pipelines used in \citetalias{EHTCIV} only produces 30 images per simulated image. This effectively increases the sampling uncertainty in the reconstructions. Nevertheless, in this section we will apply \VIDA and \ReX to the CLEAN reconstructions of the crescent with $180^\circ$ PA and the GRMHD simulation from \autoref{ssection:paperiv}. After we will repeat the analysis of the disk and double simulated EHT data from \autoref{section:asym} solely using \VIDA. We will only present results on the EHT sampling of April 11, but found similar results on the other days. 

\subsection{Ring-Like Reconstructions}

To analyze the CLEAN reconstructions we followed an identical procedure as in \autoref{sec:validation}, except we used the CLEAN image reconstructions pipeline. An example reconstruction and optimal \VIDA template for the crescent and GRMHD simulation are shown in the first two columns of \autoref{fig:clean_recon}. We see that the crescent is accurately recovered although it is generally thicker than the on-sky image, which is due to the CLEAN beam.

\begin{figure*}
    \centering
    \includegraphics[width=1.0\linewidth]{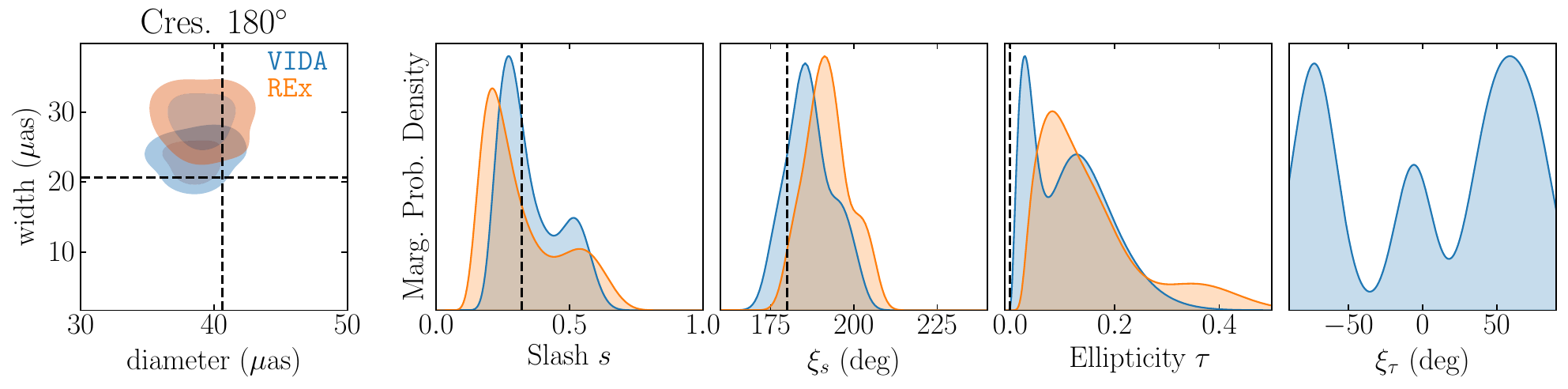}\\
    \includegraphics[width=1.0\linewidth]{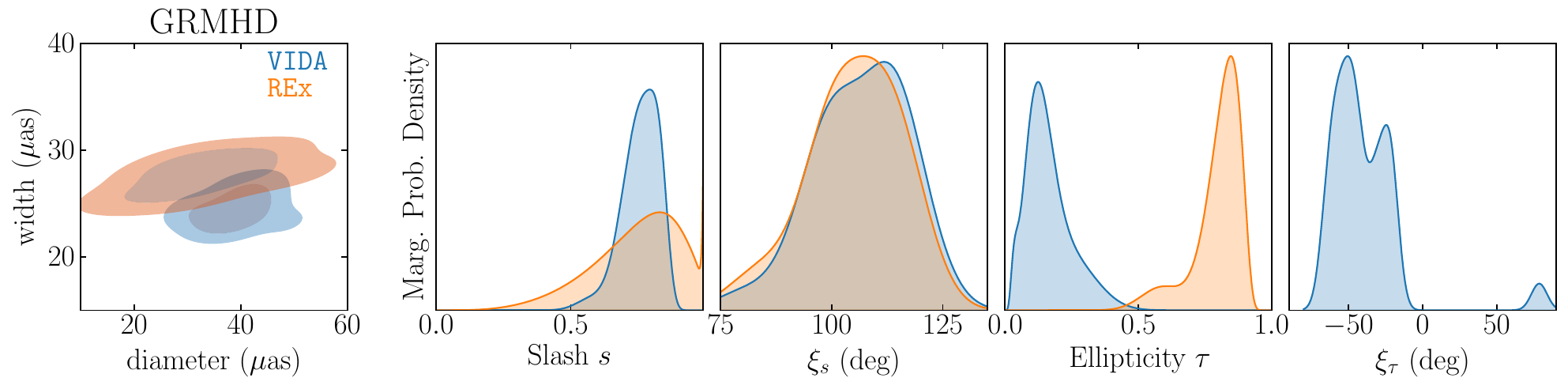}
    \caption{Marginal distributions from applying \VIDA (blue) and \ReX (orange) to the CLEAN reconstructions. The top row shows the crescent with brightness position angle $180^\circ$ east of north. Note we see similar results for the other crescent orientation. The bottom row shows the results from the GRMHD simulations. Here we see large differences in the measured parameters between \ReX and \VIDA. The origin of these differences are likely from the GRMHD simulations lacking a dominant ring in the image, breaking the underlying ring assumptions of \ReX and \VIDA.}
    \label{fig:clean_ring_marg}
\end{figure*}

Quantitatively, the marginal parameter distributions for the crescent are shown in the top panels of \autoref{fig:clean_ring_marg}. From these reconstructions we see a number of interesting differences when compared to the \ehtim results in \autoref{fig:m87geom_results}. The recovered crescent diameter is generically smaller than the \ehtim results. This difference can be explained by the larger width. That is, the measured on sky width is given by the root mean square of the CLEAN beam size and the intrinsic ring width, which is shown by the black lines in \autoref{fig:clean_ring_marg}. The measured diameter then matches the diameter of the peak brightness after blurring, i.e. \autoref{eq:diam_bias}. One interesting thing to note is that the ellipticity for the CLEAN reconstructions appears not to have the same north-south orientation bias as the \ehtim reconstructions. This suggests that the ellipticity bias we saw for the \ehtim reconstructions is due to the specific hyperparameters used in the construction of the topset. Note that while we did not show the crescent with position angle $150^\circ$, we did get similar results.

For the GRMHD results shown in the bottom row of \autoref{fig:clean_ring_marg}, the distributions between \ReX and \VIDA are in disagreement. Namely, the \ReX diameter distribution width is quite a bit larger. Additionally, the measured ellipticity is discrepant between the two feature extraction methods. This occurs because the CLEAN reconstructions of the GRMHD simulation do not show a dominant ring feature (see \autoref{fig:clean_recon}). In this case, the assumptions underlying \ReX's parameter definitions and \VIDA's ring template break down. We do note that qualitatively, \VIDA does seem to recover the visual image size and ellipticity better than \ReX. Furthermore, in this case a user would probably employ a different \VIDA template given the apparent non-ring structure. Especially since the actual on-sky image structure would not be known a priori.

\subsection{Disk and Double Reconstructions}

\begin{figure}
    \centering
    \includegraphics[width=1.0\linewidth]{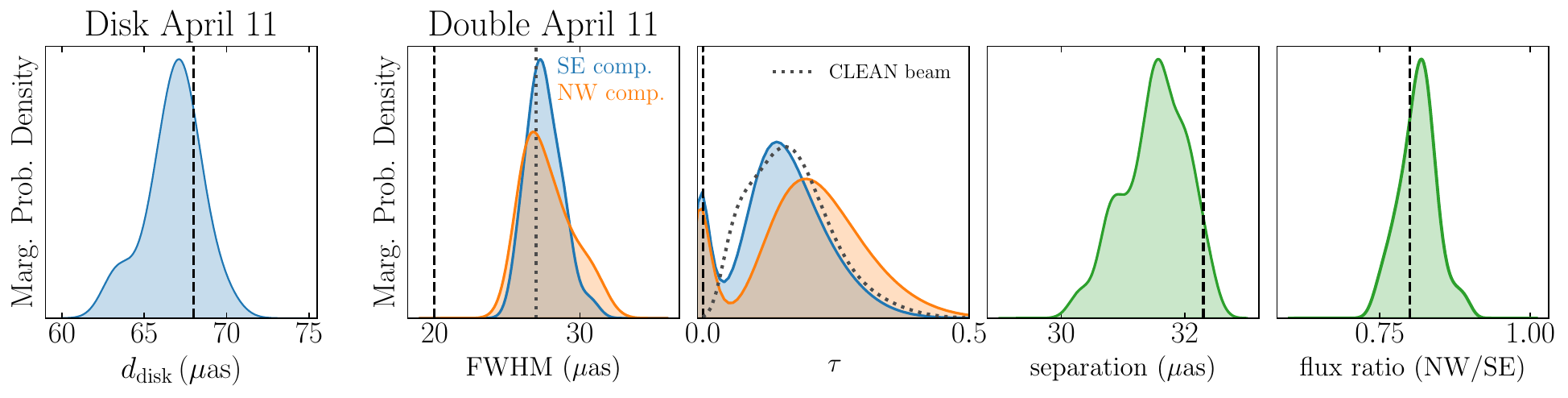}
    \caption{Marginal distribution from applying \VIDA to the disk (left panel) and double (remaining panels) simulated datasets. In this case we see that CLEAN is able to recover the diameter of the disk unlike \ehtim, which had underestimated the diameter by a few $\muas$ (see \autoref{fig:clean_recon}). For the double we see that \VIDA and the CLEAN reconstructions are able to recover the separation and flux ratio. The true ellipticity is also recovered. Note that the larger $\tau$ and fwhm for the Gaussian's follows from incorporating the contributions of the CLEAN beam (gray dotted line).}
    \label{fig:dbl_disk_marg}
\end{figure}

For the disk we used the same template as \autoref{ssec:disk}. However, for the double we only used a two-component asymmetric Gaussian template. The reason for this was the the CLEAN reconstructions only had two Gaussian components, unlike the \ehtim ones. We show example reconstructions and the optimal template for the disk and double in \autoref{fig:clean_recon}. Visually we see that the templates are reconstructing the qualitative features of both the double and disk. More quantitatively, we show the marginal parameter distributions in \autoref{fig:dbl_disk_marg}. We find near identical FWHM for both Gaussian components although both are larger than the ground truth, due to the larger CLEAN beam used. However, the rest of the parameters are consistent with the truth in all cases. For the disk, we find that unlike the \ehtim reconstructions, the disk size is included in the marginal distribution. Namely, the CLEAN reconstructions don't seem to suffer from the same size bias as the \ehtim top set. This further supports our hypothesis that the size bias is due to the \ehtim hyperparameters chosen in \citetalias{EHTCIV}.

\end{document}